\documentstyle[aasms4]{article} 
\input epsf 
\begin{document}
\title{A new method to determine Globular Cluster ages}
\author{Raul Jimenez\altaffilmark{1}}  
\affil{Royal Observatory Edinburgh, Blackford Hill EH9 3HJ, Edinburgh, UK} 

\author{Paolo Padoan\altaffilmark{2,3}}
\affil{Theoretical Astrophysics Center, Blegdamsvej 21, DK-2100 Copenhagen, Denmark}  
\affil{Dept. of Astronomy, University of Padova, Vicolo dell' Osservatorio 5, I-35122, Padova, Italy} 
\authoremail{raul@roe.ac.uk, padoan@tac.dk}

\begin{abstract}
We present a new method to compute stellar ages in Globular 
Clusters (GC) that is  ten times more precise than the 
traditional isochrone fitting procedure.  
 The method relies on accurate stellar evolutionary tracks 
and on photometry for GCs complete down to the main sequence, and it is 
based on counting number of stars in two different regions of the CMD: 
the red giant branch and the main-sequence. We have applied this 
method to the globular cluster M68 and found an age of 16.4$\pm$0.2 Gyr 
for $(m-M)_V=15.3$.
This new method reduces the error associated to the uncertainty in the 
distance modulus by a factor of two, the error due to the choice of the 
value for the mixing length parameter to almost zero and the 
error due to the colour-$T_{\rm eff}$ transformation to zero. 
\end{abstract}


\section{Introduction}

Globular clusters are among the oldest objects in the Universe and 
also the tracers of the collapse of the Galaxy, and they are very 
important cosmological probes for the age of the Universe. The determination of the age of GCs is still an open problem.  
GCs ages have been recently reviewed by Chaboyer (1995). When all 
possible random and systematic errors are taken into account, 
an error bar of 5 Gyr is associated 
to any age determination using the main sequence turnoff isochrone 
fitting method (MSTO). An alternative method has been proposed 
in order to cure some of the problems of the MSTO procedure (Jimenez 
et al. 1996), giving lower ages than the MSTO.

The main problem in the MSTO comes from the fact that the isochrone 
has to fit the position of the main sequence turnoff that is not a point on the CMD, but 
rather an extended region. The same is true for the alternative method 
developed by Iben \& Renzini (1984).

In this paper we present a new method to determine ages of GCs that 
does not rely on fitting any particular morphological feature in the CMD diagram and 
that allows us to reach a precision of 0.2 Gyr for a given distance modulus. 
The method is based on a careful computation of stellar 
evolutionary tracks and on counting stars in two different regions of the 
CMD: the red giant branch (RGB) and the main sequence (MS),  
 down to a magnitude where the sample is complete.

The paper is organized as follows. In section 2 we 
describe the theoretical stellar evolution models used in this work. In 
section 3 we present the method and compute the age for M68.  
Section 4 contains a comparison with other 
 methods. We finish with the summary and conclusions. 

\section{Theoretical stellar evolutionary models}

The set of stellar evolutionary models has been computed with 
the latest version of James MacDonald's code (JMSTAR9). The code 
incorporates the latest advances in opacities and updated 
physics, it also uses the elegant technique developed by 
Peter Eggleton to follow evolution up to the RGB tip. 
Tracks were computed for a range of masses from 0.50 $M_{\odot}$ to 
1.00 $M_{\odot}$ with a mass interval of 0.001$M_{\odot}$. This is 
achieved using an adaptive mesh grid with 3000 points. A complete 
description of the code, as well as a detailed list of the physics 
used on it, is given in Jimenez \& MacDonald (1996). 

The tracks were started from a contracting initial gas cloud in the 
Hayashi track and follow up to the RGB tip, where the Helium 
core flash occurs.
All evolutionary tracks were stopped at the Helium core flash,  
that takes place under degenerate conditions. 
The mixing length parameter was chosen from the fit to the RGB 
position (Jimenez et al. 1996); for M68 we adopted a value of 1.38. 
The metallicity for all the tracks was $Z=0.0002$ and $Y=0.24$. The 
whole grid (that covers a range of masses from 120--0.01 M$_{\odot}$) 
with several metallicities and values of $Y$ will be 
made available shortly (Jimenez, Padoan \& MacDonald 1996, in preparation).
Some of the tracks are plotted in Fig. 1.

\section{The method}
Stars of different masses evolve at a different speed along the 
CMD -- the more massive the faster -- with the effect that the number 
of stars inside a fixed luminosity bin in the main sequence decreases 
as time increases. It seems natural to use this effect as a clock to 
measure the age of the GCs, since it is as simple as counting stars 
in the CMD.

In order to predict accurately the number of stars in a theoretical 
CMD it is necessary to have a large number of evolutionary tracks in the range 
of masses observed in GCs (0.80 M$_{\odot}$ to 0.70$M_{\odot}$) for M68. 
To achieve a precision of 0.2 Gyr it is necessary to have one track every 
0.001 M$_{\odot}$ (see section IV) and therefore a high number of grid  
points (2000 or more) in the adaptive mesh of the stellar evolution code.

In addition it is necessary to have {\it complete} photometry for the 
GC to a certain magnitude along the main sequence, and very accurate 
photometry along the RGB to be able to distinguish AGB stars from 
RGB stars.

To test our method we have used photometric data obtained by 
ourselves for the GC M68 (Jimenez et al. 1996). Very 
accurate photometry was obtained in several bands (UBVRIJHK). This 
allowed us to clearly distinguish the AGB from the RGB.  
 M68 has a very low metallicity ($Z=0.0002$) and therefore is 
representative of the oldest GCs in our galaxy and its age is a 
constraint on the age of the Universe.

The first step of the method consists in comparing the theoretical and 
observational luminosity functions for the main sequence stars in order 
to determine to which magnitude the observational data are complete. 
The second step consists in sampling the 
luminosity function using only two luminosity bins, one for 
the RGB and the other for the main sequence, down to the luminosity  
where the data are complete. Note however that in 
our application of the method to M68 we hardly include in the second 
bin the top of the main sequence since our data are complete only 
to $V=19.0$.

\subsection{The theoretical luminosity function}

We first draw a set  of evolutionary tracks (luminosity vs. time) 
for a given value of the 
metallicity and helium content (the mixing length parameter has 
been fixed to 1.38 (see Jimenez et al. 1996)). We then choose an  
age, that is represented by a vertical line that intersects the tracks.
Finally, we fix luminosity values that are horizontal lines in the same 
time-luminosity diagram.

%\begin{figure}
%\centering
%\leavevmode
%\epsfxsize=1.0
%\columnwidth
%\epsfbox{globagefig1.eps}
%\caption[]{Evolutionary tracks for stars in the range of masses 0.7-0.8 M$_{\odot}$. 
%The luminosity bins and the time are the ones used for computing 
%the luminosity function shown 
%in Fig. 2. The tracks are spaced by 0.001 M$_{\odot}$.}
%\end{figure}

The track that goes through the intersection between a given luminosity 
and the time gives the mass that corresponds to that luminosity at
that time. The whole procedure is illustrated in Fig. 1.

In this way we can use stellar evolutionary tracks to determine the 
mass-luminosity ($M-L$) relation for stars of any mass and metallicity (Padoan 
\& Jimenez, in preparation).

The luminosity function is determined by using this $M-L$ relation and
by assuming a stellar initial mass function (Padoan 1995).

%\begin{figure}
%\centering
%\leavevmode
%\epsfxsize=1.0
%\columnwidth
%\epsfbox{globagefig2.eps}
%\caption[]{The theoretical luminosity function (crosses and dotted line) for 
%the estimated age of M68 is compared with the observational luminosity 
%function (diamonds). The observations are fitted remarkably well by the theory 
%down to the magnitude $V=19.0$. This indicates that the data are complete down 
%to $V=19$. The largest error bars for the observations are about the size 
%of the plotting symbols (diamonds)}
%\end{figure}

An example of a theoretical luminosity function is shown in Fig. 2 for 
the range of masses $0.71-0.77 M_{\odot}$ and for metallicity $Z=0.0002$ 
with $Y=0.24$. The observed luminosity function for M68, obtained 
excluding the AGB and HB stars, is plotted for comparison.

In the case of our data for M68, the theoretical luminosity function is 
remarkably well fitted by the theory down to a magnitude of $V=19$ (notice the linear
scale in the plot). The observational luminosity function deviates from 
the theoretical one only for magnitudes larger than $V=19$. 
Therefore we consider our data complete down to a magnitude of $V=19$.

\subsection{The age of the GC}

The second step of the method consists in sampling the 
luminosity function using only two luminosity bins, one for 
the RGB and the other for the main sequence down to the 
value where the data are complete.  

The number of stars that populate the luminosity bin in the main 
sequence is decreasing, as time increases, more rapidly than the 
number of stars in the RGB. Therefore the ratio of these two numbers 
is a function of the age of the GC. 

In Fig. 4 we show the two-bin luminosity function for different 
ages and compare it with the observational value.

%\begin{figure}
%\centering
%\leavevmode
%\epsfxsize=1.0
%\columnwidth
%\epsfbox{globagefig3.eps}
%\caption[]{This diagram shows the luminosity bins used to determine the age of M68. The 
%vertical lines are the different ages considered.  }
%\end{figure}

For a distance modulus $(m-M)_{\rm V}=15.3$ the observations are 
best fitted by an age of $16.4 \pm0.2$ Gyr.

%\begin{figure}
%\centering
%\leavevmode
%\epsfxsize=1.0
%\columnwidth
%\epsfbox{globagefig4.eps}
%\caption[]{The two-bin luminosity function. The left bin contains the RGB stars, 
%the right bin the main sequence stars down to $V=19.0$. Diamonds connected by dotted 
%lines are the theoretical two-bin luminosity functions for different ages spaced 
%by 0.5 Gyr. The continuous line is the observational value. We have 
%plotted the 2\% error bar due to the uncertainty in counting stars, 
%which corresponds to and uncertainty in the age of 0.2 Gyr.}
%\end{figure}

The age determination depends on the assumed distance modulus. In Table
1 ages are given for different distance moduli. As expected the cluster
appears older when it is assumed to be closer (smaller distance modulus).
Jimenez et al (1996) have determined the distance modulus with high
precision by fitting the luminosity function of the RGB with theoretical 
luminosity functions (from stellar evolutionary tracks). Their result is
$(m-M)_{\rm V}=15.3 \pm 0.1$. Therefore our best estimate of the age of 
the globular cluster M68 is $16.4 \pm 0.2$ Gyr for the assumed distance 
modulus. If the uncertainty in the distance modulus is considered, the
uncertainty in the age is $\pm 1.5$ Gyr. 

%\begin{table}
%\begin{tabular}{cc}
%$(m-M)_V$ & Age (Gyr) \\
%\hline
%15.5 & $14.5 \pm 0.2$ \\
%15.4 & $15.0 \pm 0.2$ \\
%15.3 & $16.4 \pm 0.2$ \\
%15.2 & $17.5 \pm 0.2$ \\
%15.1 & $18.8 \pm 0.2$ \\
%15.0 & $20.0 \pm 0.2$ \\
%\hline
%\end{tabular}
%\caption{Age determination for different distance moduli}
%\end{table}

\subsection{Accuracy of the method}

To estimate the error due to counting stars we proceeded in the following way. 
Several frames of the same clusters taken during a period of 15 nights and 
different seeing conditions were analysed to count the total number of 
star in the frame and the number of stars per bin of luminosity. 
This allowed us to 
estimate the error that comes from crowding and  choice of the point 
spread function. For this purpose we used 20 frames. The difference in the 
total number of stars from frame to frame was not bigger than 2\%, and 
the same applies when the stars were counted in the bins used in the method.
This corresponds to an error of 0.15 Gyr.

We have also checked the effect of the IMF on the age determination. In this 
work we use a power law IMF with exponent 2.0 (where Salpeter is 2.35). 
A steeper IMF with exponent 3.0 affects the age only slightly  in the sense 
of making the GC older by only 0.1 Gyr. Therefore we conclude that for reasonable 
IMF slopes the error related to the IMF is about 0.1 Gyr.

It is interesting to point out how stable the stellar evolution code is. 
Stellar tracks spaced by 0.001$M_{\odot}$ are clearly defined in the 
time-luminosity diagram (see fig. 1). We tried to understand how much the 
position of the tracks in such a diagram could change due to different 
initial conditions in the starting protostellar cloud, and to round-off
errors in different computers. The result of computations with different 
initial conditions and with different machines has shown that the  
computed tracks are very stable, in the sense that they occupy the same 
position in the time-luminosity diagram with a precision such that two 
tracks spaced by only 0.0001$M_{\odot}$ can be distinguished, when the
stellar evolution code is run with 3000 mesh points. Therefore the
uncertainty in the theoretical determination of stellar masses does
not affect the method at all.  

As stated by Chaboyer (1995) the main uncertainty in the MSTO method is the 
choice of the value of the mixing length parameter ($\alpha$). This gives  
uncertainties in the age as large as 10\%. In our method the value 
of $\alpha$ does not affect the age determination, since we use only  
tracks in the time-luminosity diagram that look almost identical even for very different values of $\alpha$ (Jimenez et al 1996). Therefore the mixing 
length parameter is not a source of error for us as in the MSTO method.  

Another source of error in the MSTO is the transformation between 
colour and $T_{\rm eff}$. Chaboyer (1995) gives an estimate of 5\%. 
In our method the error due to this is zero since no colour transformation 
is necessary to compute the luminosity function.

Finally we comment briefly about how the uncertainty in the value 
of the distance modulus affects our age determination. Again, 
Chaboyer (1995) computes an error of 25\% on the age due to uncertainties 
in the distance modulus value for the MSTO. It transpires from Table 1 
that in our method that uncertainty has been reduced to 15\%.

The total error in our age determination is estimated to be 0.2 Gyr that is 
the sum of the uncertainty in the number of stars per bin and in the IMF slope. 
In addition, an error of 1.5 Gyr should be added when the distance modulus 
is not known better than 0.25$^m$. It should be stressed that the same 
uncertainty (0.25$^m$) gives in our method an uncertainty in the age 
of only 15\%, but 25\% in the MSTO.

In the case of MSTO if the distance modulus, $\alpha$, colour-$T_{\rm eff}$  
transformation and chemical composition are fixed to a certain value, the uncertainty in the 
age is 10\%, while it is only 2\% in the present method.  

\section{Discussion}

The investigation of GCs ages requires the discussion of two
basic problems:
\begin{itemize}
\item The determination of the stellar absolute luminosity from the 
observed stellar magnitudes, that is the problem of measuring 
distances accurately.

\item The uncertainties in stellar evolution theory that translate into
uncertainties in the prediction of stellar ages.  
\end{itemize}

A third problem arises in the age determination method based on
isochrone fitting. Namely this method presents the problem of 
fitting the position of the MSTO that is not a point on the CMD,
but rather an extended region. In fact the position of the MSTO is 
very sensitive to the assume mixing length parameter and colour calibration. The same is true for the alternative 
method developed by Iben \& Renzini (1984).

The method developed in the present work is also affected by the
uncertainties in the estimated distance of the globular cluster and
in the stellar evolution theory. Nevertheless it improves
considerably on the previous ones (by a factor 10!) because it does not rely on
fitting any particular morphological feature in the CMD, and does not depend 
at all on mixing length parameter and colour calibration. In fact
it has been shown in the paper that an uncertainty of only $0.2$ Gyr
is achieved, for a given distance modulus, just by counting stars 
on the CMD, as long as the stellar counts are stopped at a magnitude
where the data are known to be complete. 

As far as the stellar evolution theory is concerned, two are the 
most important uncertainties:

\begin{itemize}

\item The enhancement of $\alpha$ elements in GCs (Pagel \& Tautvaisiene 1995).  
How to handle them in stellar evolution theory is still an open problem 
(Vandenberg 1992, Salaris, Chieffi \& Straniero 1993).

\item The helium settling in the radiative core that can reduce the amount 
of H and therefore shorten stellar ages.

\end{itemize}

The stellar evolution models used in this work do not include any of these 
effects. A simple solar-scaled composition has been used, and no He 
diffusion has been taken into account. Nevertheless, these uncertainties do 
not invalidate our procedure. If $\alpha$ elements and He diffusion 
affect significantly the stellar ages, our method would give an age 
estimate for the globular cluster shortened by 20--30\%, that is an 
age in agreement with previous works (Chaboyer, Sarajedini \& Demarque 1992, 
Jimenez et al. 1996).

\section{Summary and conclusions}

We have developed a new method to determine the ages of GCs. Using 
theoretical evolutionary tracks we have predicted the relative number of stars
in the main sequence and in the RGB, as a function of age and distance  
modulus. 

The dependence of the age on the distance modulus is twice smaller than 
what is found using the traditional isochrone fitting method, but the accuracy 
of the age determination for a given distance modulus is ten times higher,
 because the present method is based just on counting stars 
in different luminosity bins and therefore does not have troubles 
with fitting the morphology of the MSTO (mixing length parameter and colour calibration).

In Table 2 we show a comparison of the errors involved in computing GCs ages using the MSTO and our method.

We have applied this method to the old halo GC M68 and found an age 
of $16.4 \pm 0.2$ Gyr, if the distance modulus $(m-M)_{\rm V}=15.3$ determined 
by Jimenez et al. (1996), is used.

This value is in good agreement with previous age determinations found 
by Chaboyer, Sarajedini, Demarque (1992), using isochrone fitting, and Jimenez et al. 
(1996), using the HB morphology technique.

\acknowledgements
This work has been partly supported by the Danish National Research Foundation 
through its support for the establishment of the Theoretical Astrophysics Center.
RJ thanks the Theoretical Astrophysics Center in Copenhagen where part 
of this work was carried out. 
 PP enjoyed the hospitality of the Royal Observatory in Edinburgh were part of this 
work was carried out.  
We thank James Macdonald for kindly providing us the latest version of 
his stellar evolution code (JMSTAR9).

\newpage

\begin{table}
\begin{tabular}{cc}
$(m-M)_V$ & Age (Gyr) \\
\hline
15.5 & $14.5 \pm 0.2$ \\
15.4 & $15.0 \pm 0.2$ \\
15.3 & $16.4 \pm 0.2$ \\
15.2 & $17.5 \pm 0.2$ \\
15.1 & $18.8 \pm 0.2$ \\
15.0 & $20.0 \pm 0.2$ \\
\hline
\end{tabular}
\caption{Age determination for different distance moduli}
\end{table}

\newpage

\begin{table}
\begin{tabular}{ccc}
 & MSTO & This work \\
\hline
distance modulus & 25\% & 15\% \\
mixing length & 10\% & 0\% \\
colour--$T_{\rm eff}$ & 5\% & 0\% \\
He diffusion & 7\% & 7\% \\
$\alpha$-elements & 10\% & 10\% \\
\hline
\end{tabular}
\caption{The values of the errors associated with different 
uncertainties when computing GCs ages. The MSTO and our method have 
been compared. Notice how the influence of this uncertainties 
in our method are smaller than in the MSTO. An accuracy in the 
age determination of 5\% can be achieved with our method.}
\end{table}

\newpage

\begin{figure}
\centering
\leavevmode
\epsfxsize=1.0
\columnwidth
\epsfbox{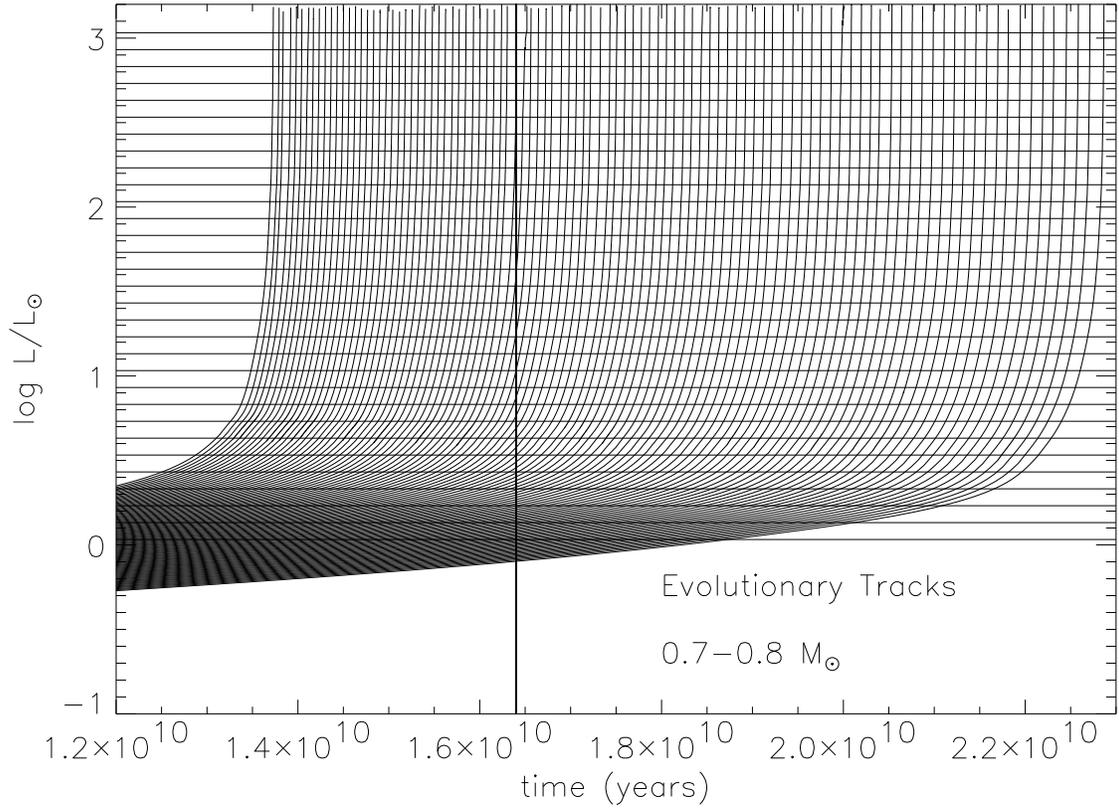}
\caption[]{Evolutionary tracks for stars in the range of masses 0.7-0.8 M$_{\odot}$. 
The luminosity bins and the time are the ones used for computing 
the luminosity function shown 
in Fig. 2. The tracks are spaced by 0.001 M$_{\odot}$.}
\end{figure}

\newpage

\begin{figure}
\centering
\leavevmode
\epsfxsize=1.0
\columnwidth
\epsfbox{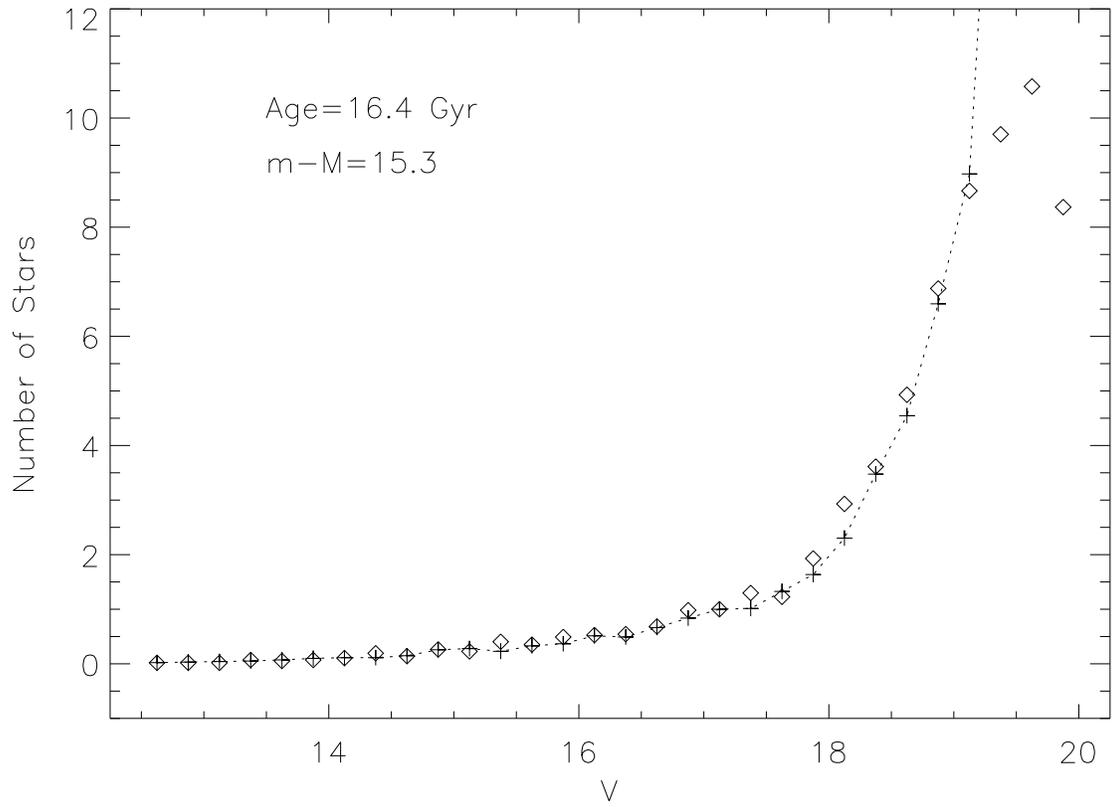}
\caption[]{The theoretical luminosity function (crosses and dotted line) for 
the estimated age of M68 is compared with the observational luminosity 
function (diamonds). The observations are fitted remarkably well by the theory 
down to the magnitude $V=19.0$. This indicates that the data are complete down 
to $V=19$. The largest error bars for the observations are about the size 
of the plotting symbols (diamonds)}
\end{figure}

\newpage

\begin{figure}
\centering
\leavevmode
\epsfxsize=1.0
\columnwidth
\epsfbox{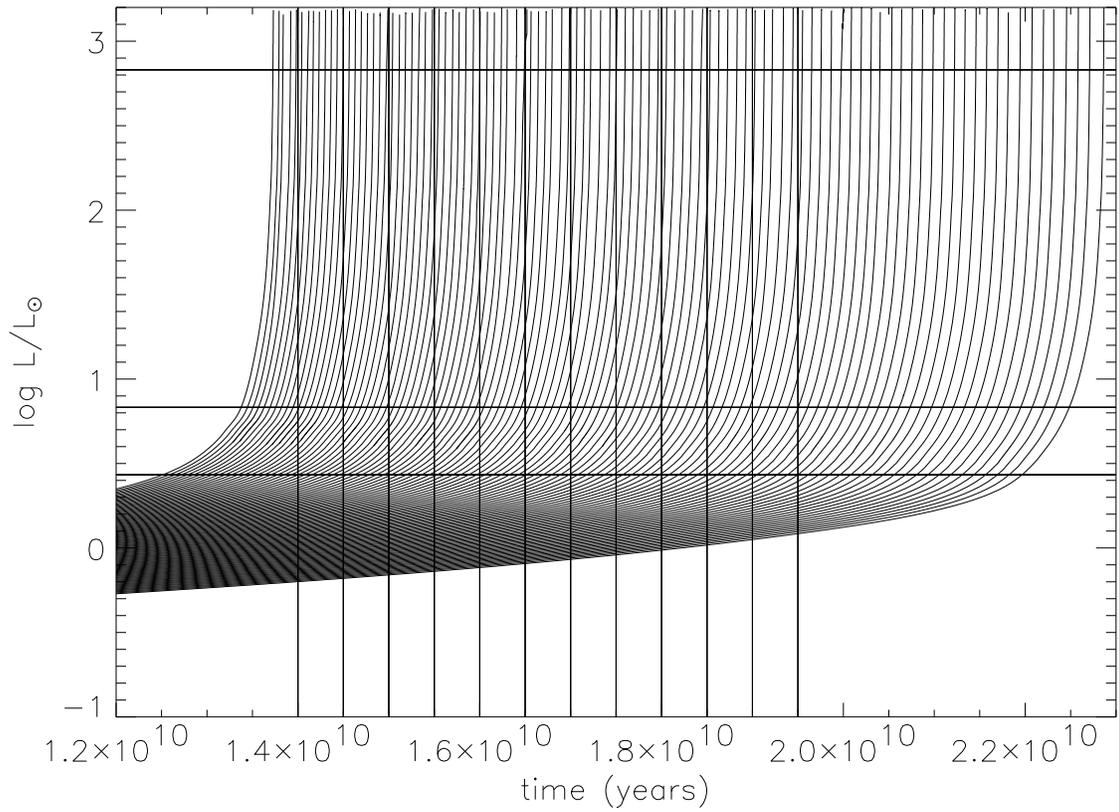}
\caption[]{This diagram shows the luminosity bins used to determine the age of M68. The 
vertical lines are the different ages considered.  }
\end{figure}

\newpage

\begin{figure}
\centering
\leavevmode
\epsfxsize=1.0
\columnwidth
\epsfbox{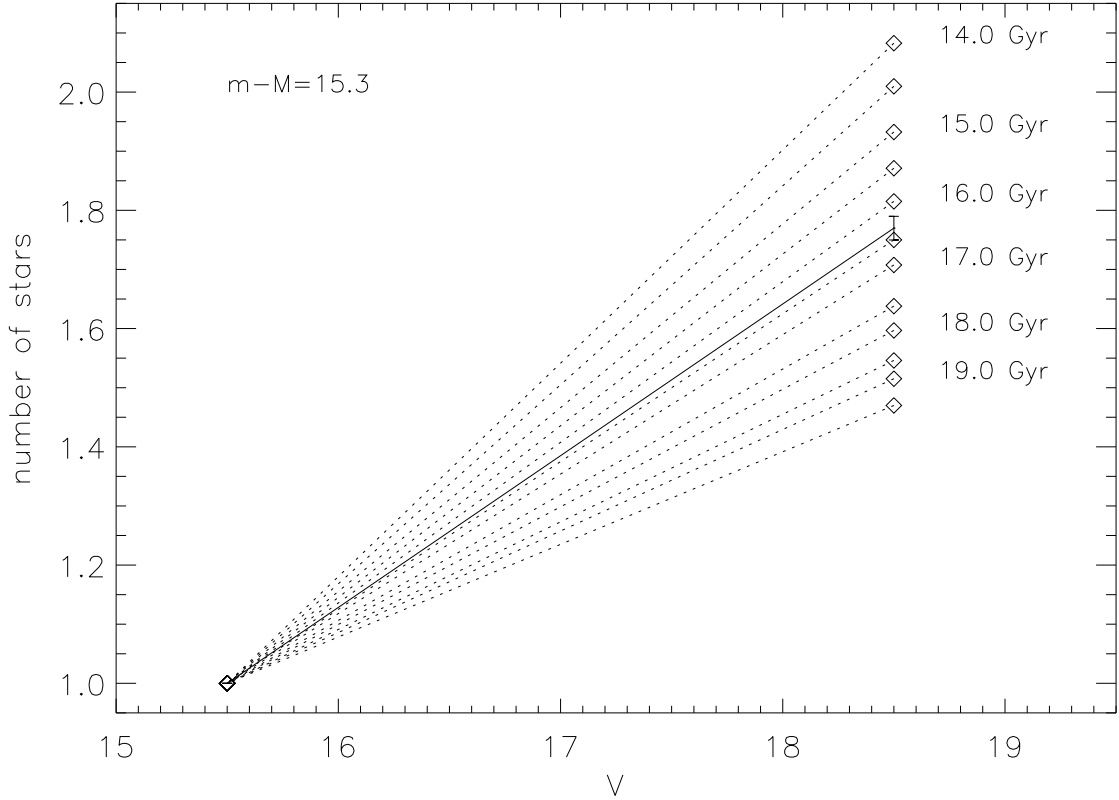}
\caption[]{The two-bin luminosity function. The left bin contains the RGB stars, 
the right bin the main sequence stars down to $V=19.0$. Diamonds connected by dotted 
lines are the theoretical two-bin luminosity functions for different ages spaced 
by 0.5 Gyr. The continuous line is the observational value. We have 
plotted the 1\% error bar due to the uncertainty in counting stars, 
which corresponds to and uncertainty in the age of 0.2 Gyr.}
\end{figure}

\end{document}